\documentstyle[12pt]{article}
\begin{document}
\begin{titlepage}
\vspace*{-0.5cm}
\begin{flushright}
DTP/94/10\\
June 1994\\
\end{flushright}
                               
\vskip 1.cm
\begin{center}                                                                  
{\Large\bf Anomalous quartic couplings in\\
\vskip .3cm
 $    W^+W^- \gamma $ production at $    e^+e^- $ colliders }
\vskip 1.cm
{\large Ghadir~Abu~Leil}
\vskip .2cm

and
\vskip .2cm

{\large W.J.~Stirling}
\vskip .2cm
{\it Department of Physics, University of Durham \\
Durham DH1 3LE, England }\\
\vskip .4cm

\vskip   .4cm

\vskip .8cm                                                                    
\end{center}                                                                    
\begin{abstract}
 We study the process  $e^+e^- \rightarrow  W^+W^- \gamma$
 at high-energy  $e^+ e^-$ colliders to
 investigate the effect of
genuine quartic  $W^+W^-\gamma\gamma$ and $W^+W^- Z\gamma$ anomalous
couplings on the cross section. Deviations from the
Standard Model predictions are quantified. We show how bounds on the
anomalous couplings can be improved by choosing specific initial
state helicity combinations. The dependence of the anomalous contributions
on the collider energy is studied.
\end{abstract} 
\vfill
\end{titlepage}
\newpage

\section{Introduction} 
 In the Standard Model (SM) of electroweak interactions, the
SU$_{L}$(2)$\times$ U$_{Y}$(1) non-abelian nature of the gauge symmetry
relates  the trilinear and quadrilinear
vertices to the universal  SU(2) gauge coupling,
$g_w$. At tree level there are only two
trilinear vertices, $W^+W^- \gamma $ and  $W^+W^- Z $, and four
quartic vertices $ W^+W^-
\gamma \gamma$, $W^+W^- \gamma Z$, $W^+W^- Z Z$ and $W^+W^-W^+W^-$.
Only recently have
experiments begun to test these vertices directly. At the CERN and FNAL
$p \bar p$ colliders, a handful of $W^{\pm}\gamma$ events have been
used to place
limits on the anomalous $W^+W^- \gamma$ and $W^+W^- Z$ trilinear
 couplings \cite{UA2,CDF,D0}.
The LEP II $e^+e^-$ collider will also test the trilinear vertices
through the total $W^+W^-$ cross section  \cite{GREENBOOK}.
 However, independent tests
of the {\it quartic} couplings require more complicated processes.
One of the most accessible in the short term is the process
$e^+e^- \rightarrow W^+W^- \gamma$, and it is this which provides 
the focus of the present study.

 Studying the gauge boson
interactions (of the $W$ boson  in particular) will help our
 understanding  of  the  mechanism of spontaneous symmetry breaking. Since the
gauge-boson  self interaction originates in  the
non-abelian kinetic term of the Lagrangian, it is
directly  related to the Goldstone modes and the Higgs particle.
The quartic couplings in particular will  provide a way of testing
the Higgs mechanism, either verifying the
local gauge invariance or  signaling
 the existence of new physics beyond the Standard Model.  A review
 of the importance of quartic couplings in probing new physics
 can be found in Ref.~\cite{REDBOOK}.

There is  an important distinction  between anomalous trilinear and
genuine anomalous quartic couplings, {\it i.e.} those which give no
contribution to the trilinear vertices \cite{BOUDJEMA}. Whereas
the trilinear couplings involving $W$'s  are  essentially form factors
where massive fields are integrated out at the one-loop level, the
anomalous  quartic couplings
 are contact interactions  --  manifestations of the exchange of
heavy particles. One can therefore imagine  a theory in which
the trilinear couplings have  their Standard Model  values, but  the quartic
couplings are  modified by any number of independent anomalous
contact interactions.

In this paper we study the effect of anomalous quartic couplings
in the process  $e^+e^- \rightarrow W^+W^- \gamma $  at high energy.
Our work builds on
and extends the analysis of \cite{REDBOOK,BOUDJEMA}, in that we investigate
the collider energy and polarization dependence of the anomalous effects. 
In Section $2$ we   discuss the contributions  of the anomalous
operators in the context  of $W^+W^- \gamma$ production and in Section $3$
the numerical results are presented. 

\section{The interaction Lagrangian}
 In this section we discuss the lowest dimension operators which lead to
genuine quartic couplings.
These operators must of course have the proper Lorentz structure, and
should
also respect the custodial SU(2) symmetry  in order
 to evade experimental bounds on the $\rho$
parameter \cite{EBOLI}. The phenomenological Lagrangian should  also 
respect the full U(1)
gauge invariance, as at least one of the fields is a  photon.
For simplicity,
we restrict the study to $C$-  and $P$-conserving operators. The lowest
dimension operators that satisfy the above constraints
 are  of dimension 6, since the U(1)$_{em}$ symmetry
 requires derivatives \cite{BOUDJEMA}.
These operators\footnote{Note   that the operator ${\cal L}^0$ 
can be parametrized by the exchange
of a neutral scalar particle.}
are: 
\begin{eqnarray}
 {\cal L}^0 &=&-\frac{{\pi}{\alpha}}{4 \Lambda^2}a_{0} F_{\alpha \beta}
F^{\alpha \beta} (\vec{W}_\mu \cdot \vec{W}^{\mu}) \\
{\cal L}^c &=&-\frac{{\pi}{\alpha}}{4 \Lambda^2}a_{c} F_{\alpha \mu}
F^{\alpha \nu} (\vec{W}^{\mu}\cdot \vec{W}_\nu) \\
{\cal L}^n & = &i \frac{{\pi}{\alpha}}{4 \Lambda^2}a_{n} \epsilon_{ijk}
W^{(i)}_{\mu \alpha} W_{\nu}^{(j)}  W^{(k)\alpha}F^{\mu \nu}
\end {eqnarray}
   where $\vec{W}_{\mu}$ is an SU(2) triplet, and $F^{\mu \nu}$ 
and  $\vec{W}^{\mu\nu}$ are the U(1)$_{em}$ and the SU(2) field strengths
respectively.
The parameter  $\Lambda$ is an unknown `new-physics'  scale which, following
convention,  we take to be  $M_W$.

The physical  Lagrangians are obtained when the above
are written in terms of the physical fields $W^+$, $W^-$
 and $Z^0= W^3 \cos\theta_w$. The physical basis  for ${\cal L}^0$
and ${\cal L}^c$ is obtained by the substitution \cite{BOUDJEMA}
\begin{equation}
\vec{W}_\mu \cdot \vec{W}_\nu \rightarrow \ 2
W^+_{\mu}W^-_{\nu} + \frac{1}{\cos\theta_w^2} Z_{\mu} Z_{\nu}
\end{equation}
 while  the physical basis for  the part of ${\cal L}^n$ which gives
 rise to quartic couplings is
\begin {eqnarray}
\vec{W}_{\mu\alpha} \cdot (\vec{W}_{\nu} \times \vec{W}^\alpha)& \rightarrow &
\frac{i}{\cos\theta_w} \left[ 
(\partial_{\mu} W_{\alpha}^+ -\partial_{\alpha} W_{\mu}^+)
(Z_\nu W^{-\alpha}-Z^{\alpha}W_{\nu}^-)   \right. \nonumber\\
&&+(\partial_{\mu} W_{\alpha}^--\partial_{\alpha} W_{\mu}^-)
(Z^{\alpha} W_{\nu}^+ - Z_{\nu}W^{\alpha +} ) \nonumber\\
&& \left. +(\partial_{\mu} Z_{\alpha} -\partial_{\alpha} Z_{\mu})
(W_\nu^- W^{+\alpha}-W^{+}_{\nu}W^{-\alpha}) \right]    
\end{eqnarray}
The effective Lagrangians ${\cal L}^0$ and ${\cal L}^c$ give rise to
an anomalous $W^+W^- \gamma \gamma$ coupling, whereas ${\cal L}^n$
gives rise to
an anomalous $W^+W^- Z \gamma$ coupling. The  corresponding Feynman rules
are listed in Ref.~\cite{EBOLI}.

\section{Numerical results and conclusions }
  
 We begin  this section  by analysing the effect of the anomalous
couplings $a_{0}$, $a_{c}$ and $a_{n}$ on the total  $W^+W^- \gamma$
production  cross section at
a $500 \ \rm GeV$ $e^+e^-$ collider \cite{REDBOOK}. The
anomalous cross sections are quadratic functions of the parameters
$a_{0}$, $a_{c}$ and  $a_{n}$. Fig.~1 shows the total
 cross-sections  with one parameter
being different from zero at any one time. 
In order to avoid collinear singularities caused by the
massless photon the following rapidity and energy cuts are implemented
\begin{equation}
 \vert \eta_\gamma \vert \leq 2\; , \qquad E_\gamma \geq 20\ \rm GeV
\end{equation}
In addition, all the initial and final particles are
 separated by at least $15^\circ$. Other parameter values are
 $M_W=80 \ \rm GeV/c^2$,
 $\sin^{2}\theta_W=0.23$ and  $\Gamma_Z = 2.55\ \rm GeV$.
 With these parameters, the 
Standard Model total cross section
($a_0 = a_c = a_n = 0$) is $123.4\ \rm fb$, which corresponds to 
a total of $N(W^+W^- \gamma)= 1234$  events for an integrated luminosity of 
${\cal L}= 10\ \rm fb^{-1}$. 
The two horizontal lines in Fig.~1 correspond to a $ \pm 3 \sigma$
 statistical variation of the Standard Model result, {\it i.e.} 
\begin {equation}
\delta \sigma_{SM}=  \pm 3\; \sqrt {\frac
{\sigma_{SM}}{ \cal {L}}}
\end{equation}
 where the integrated luminosity is again taken to be ${\cal L} = 10\ 
\rm fb^{-1}$.
The $ \pm 3 \sigma$ band corresponds to the following variation
in the anomalous couplings:
\begin {eqnarray}
-0.64 \leq &a_{0}& \leq 0.42 \nonumber \\
-1.38 \leq &a_{c}& \leq 0.65 \nonumber \\
-3.9 \leq &a_{n}& \leq 4.25 \; ,
\end{eqnarray}
indicating that the sensitivity is greatest for the $a_0$ parameter
and least for the $a_n$ parameter.

We next investigate the dependence of the cross sections on the photon
energy.
Fig.~2 shows the $E_\gamma$  distribution at $\sqrt{s} = 500\ \rm GeV$,
with the same cuts and parameters as before. Fig.~2(a) shows 
the distributions for $a_0 = 0$ (Standard Model, solid line), 
 $a_0 = \pm 1$ (dashed lines) and $a_0 = 0.42$
(the $3\sigma$ value, dotted line), the other anomalous couplings
 being set to zero.
Evidently the bulk of the sensitivity  comes from the hard photon end
of the spectrum. This is not unexpected, since the additional contributions
do not give rise to infra-red singularities as $E_\gamma \rightarrow 0$. 
Similar remarks apply to the other parameters. Figs.~2(b) and (c)
show the effect on the photon energy distribution of varying
$a_c$ and $a_n$ respectively.

To try and improve the sensitivity to the anomalous couplings, 
we consider next the helicity decomposition of the cross section.
The amplitude for $e^+e^-\rightarrow W^+W^-$ contains two different
types of contribution:
$s$-channel $Z,\gamma$ exchange and $t$-channel neutrino exchange.
The anomalous quartic coupling 
contributions to $W^+W^-\gamma$ production, however, only receive
contributions from the former,  {\it i.e.} $e^+e^-\rightarrow
Z^*,\gamma^* \rightarrow W^+W^-\gamma$.
It follows that the effects will be largest in the {\it positive helicity}
initial-state configuration, $\lambda_{e^-}\lambda_{e^+} = +1$, since
this receives no contribution from the `Standard Model background'
neutrino-exchange diagrams.
Fig.~3 shows the distribution
$ d\sigma^{\pm}(a_i)/dE_\gamma 
\ (i=0,c)$  at $500 \ \rm GeV$
for (a) the positive helicity $\lambda_{e^-}\lambda_{e^+} = +1$ cross section 
($\sigma^+$)
 and (b) the negative helicity $\lambda_{e^-}\lambda_{e^+} = -1$ cross
  section ($\sigma^-$).
For the same variation in the $a_i$, the effect is indeed much larger
in the former. 

Unfortunately,  at these energies the positive helicity 
cross section is in absolute
terms much smaller than the negative helicity cross section. This is 
illustrated
in Fig.~4, which shows the spin decomposition of the total $W^+W^-\gamma$
Standard Model cross section as a function of $E_{\rm beam}$. There is 
a difference of some two orders of magnitude  between  $\sigma^-$ and 
 $\sigma^+$. 
 
 Finally, we address the question of whether there is any possibility
 of seeing an effect in $W^+W^-\gamma$ production at lower $e^+e^-$ collider
  energies. We consider variations in $a_0$ only -- similar
  remarks apply to the other couplings. The problem at lower
 energies is that phase space restricts the photon to be soft, which is where
 the sensitivity to the anomalous couplings is least.
This is illustrated in  Fig.~5, which shows the ratio 
 of $\sigma$, $\sigma^+$  and $\sigma^-$ for $a_{0}=1$ 
 to that of the corresponding Standard Model cross section, as a function
 of $E_{\rm beam}$, with the same photon cuts as before.
Below $E_{\rm beam} =  150  \ \rm GeV$ the effects are negligible.
The increased sensitivity to $a_0$ in  $\sigma^+$ is partially offset
by the much smaller cross section in this channel. Taking
${\cal L} = 10\ \rm fb^{-1}$ for both the positive and negative
helicity channels, we calculate from Fig.~5  that  at 
$500 \ \rm GeV$, $a_0 = 1$
gives a  $ 7.5 \sigma$ increase of $\sigma^-$ 
and a $ 47 \sigma$ increase of $\sigma^+$. The corresponding numbers
for $300 \ \rm GeV$  collisions are  $ 0.4 \sigma$ 
and  $ 1.0 \sigma$ respectively.

Of course we do expect to obtain a handful of  $W^+W^-\gamma$ events
even at LEP~II, and from  these it will be possible to 
derive very crude limits on the anomalous quartic couplings.
Fig.~6 shows the total $W^+W^- \gamma$ cross sections
for $E_\gamma  > 20 \ \rm GeV$,  $\vert\eta_\gamma\vert < 2$ photons
in $e^+e^-$ collisions at $200 \ \rm GeV$, as a 
function of  $a_0$ and $a_c$.\footnote{the dependence on $a_n$ is negligible
at this energy}
Again, the dependence is quadratic. Note the vastly expanded horizontal
scale compared to Fig.~1.

In conclusion,  quartic
couplings  can provide  a window on new physics beyond the
Standard Model. We have quantified the effect of various types
of anomalous operators on the $W^+W^-\gamma$ production cross section
in $e^+e^-$ collisions. The effects are largest in the positive
helicity cross section, although this represents   only a small fraction
of the total cross section. This type of  physics is best suited
to high energy colliders -- there is an enormous increase in sensitivity in
going from $\sqrt{s} = 300\ \rm GeV$ to $\sqrt{s} = 500\ \rm GeV$  -- 
although some crude limits should
be possible even from a handful of events at LEP~II.

\section*{Acknowledgments}

This work was supported in part by an ICSC World Laboratory Studentship 
and a University of Durham Research Studentship.

\newpage

\noindent{{\Large\bf Figure Captions}}                                         
\begin{itemize}
\item [{[1]}]
The total cross section  for the process $e^+e^- \rightarrow W^+W^- \gamma$ 
at $\sqrt{s} = 500\ \rm GeV$ as a
function of the anomalous couplings $a_{0}$, $a_{c}$ and  $a_{n}$. The
$ \pm 3 \sigma$ variation 
about the SM  cross-section is indicated by the horizontal lines.

\item [{[2]}]
The  photon energy distribution $d\sigma/dE_\gamma$ 
 for different values of  (a) the $a_{0}$ parameter,
(b) the $a_{c}$ parameter, and (c) the $a_{n}$ parameter, at
$\sqrt{s} = 500\ \rm GeV$.

 \item [{[3]}]
The (a) positive-helicity  and (b) negative-helicity cross sections
 $d\sigma^{\pm}(a_i)/dE_\gamma$
as a function of the photon energy, at $\sqrt{s} = 500\ \rm GeV$.

\item [{[4]}] 
The positive- and negative-helicity contributions to the 
Standard Model $e^+e^- \rightarrow W^+W^- \gamma$ cross section as a
 function of 
the beam energy.

\item [{[5]}]
 The ratio of the total, negative-helicity and positive-helicity
cross sections for $a_{0}=1$ to those of the Standard Model ($a_{0}=0$),
as a function of the beam energy.

\item [{[6]}]
The dependence of the total $e^+e^- \rightarrow W^+W^- \gamma$ cross 
section  on the
 anomalous couplings $a_{0}$ and  $a_{c}$ at LEP~II 
($\sqrt{s} = 200\ \rm GeV$).
 
\end{itemize}

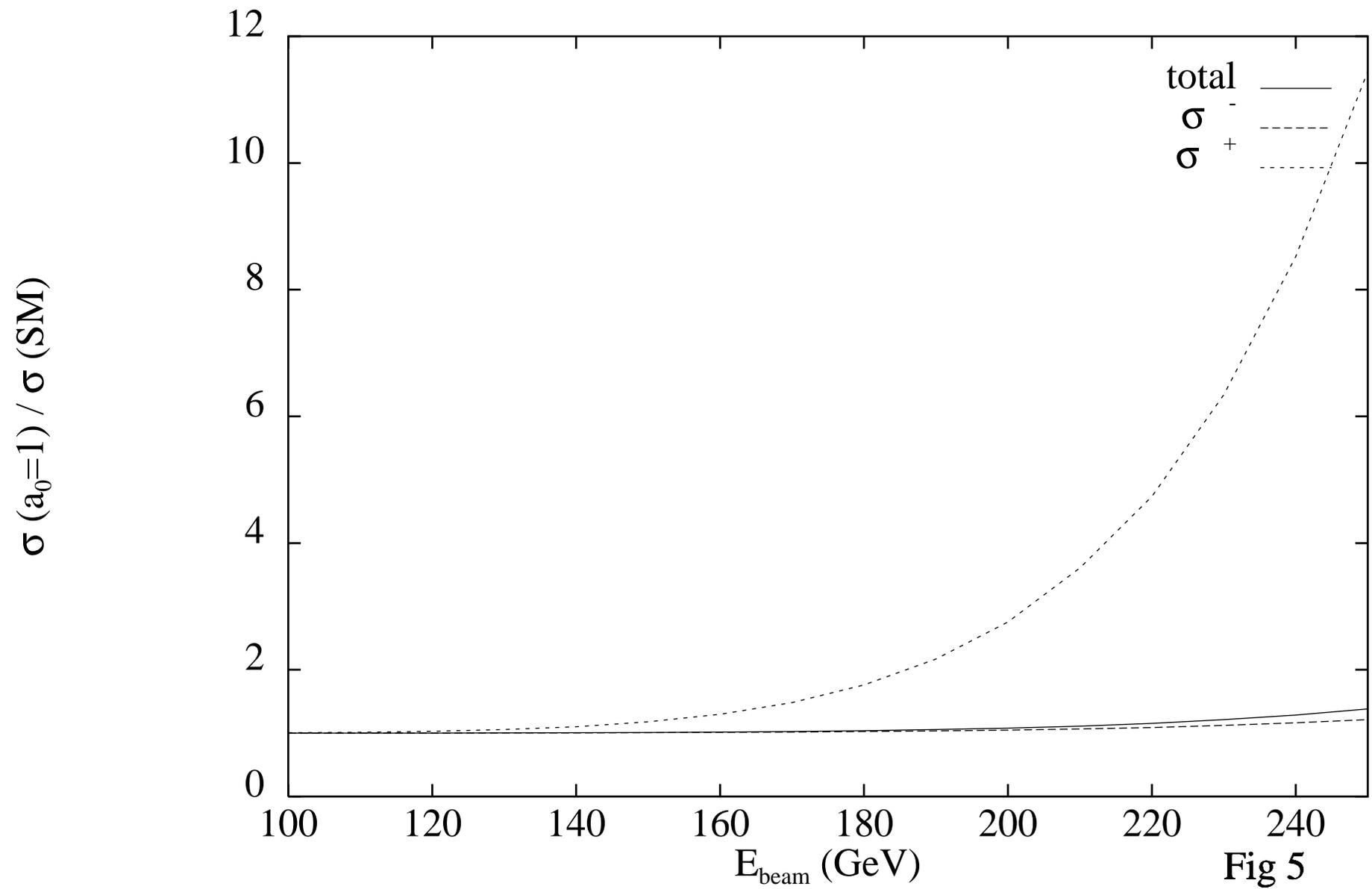
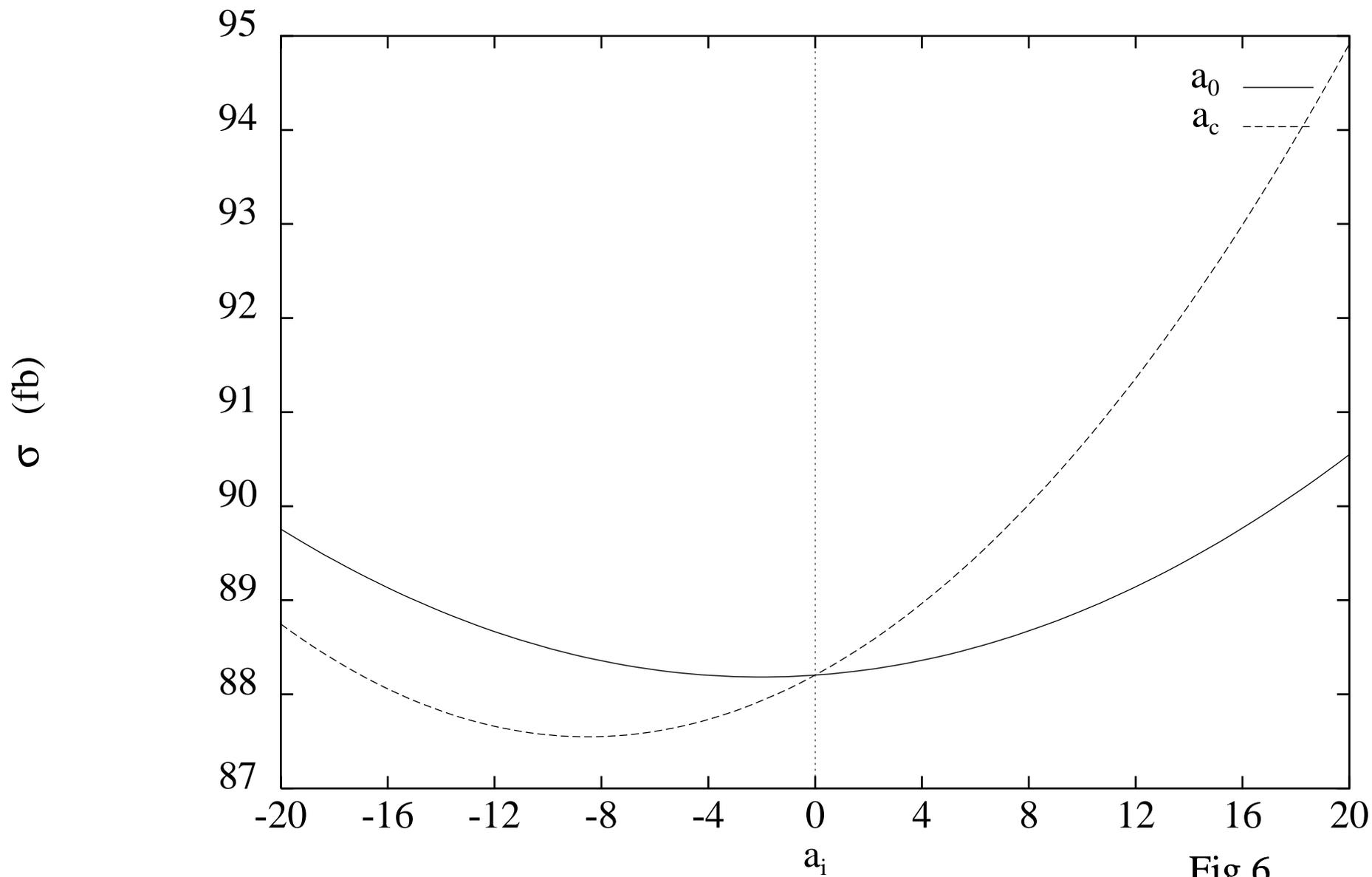
\end{document}